\documentstyle[epsf,twocolumn,aps]{revtex}
\newcommand{\eq}{\begin{equation}}
\newcommand{\ee}{\end{equation}}
\newcommand{\eqa}{\begin{eqnarray}}
\newcommand{\eea}{\end{eqnarray}}
\newcommand{\bt}{{\bf T}}
\newcommand{\btdg}{{{\bf T}^\dagger}}
\newcommand{\bs}{{\bf S}}
\newcommand{\btt}{{\bf t}}
\newcommand{\br}{{\bf r}}
\newcommand{\bunity}{{\bf 1}}
\newcommand{\bp}{{\bf P}}
\newcommand{\bu}{{\bf U}}
\newcommand{\cj}{{\cal J}}
\newcommand{\tr}{{\rm Tr}}
\newcommand{\gohm}{{g_{\rm ohm}}}

\newcommand{\pprl}{Phys. Rev. Lett. \ } 
\newcommand{\pprb}{Phys. Rev. {B}} 
\begin{document}
\twocolumn[
\hsize\textwidth\columnwidth\hsize\csname@twocolumnfalse\endcsname
\draft

\title{Conductances, Conductance Fluctuations, and Level Statistics
on the Surface of Multilayer Quantum Hall States}

\author{Vasiliki Plerou and Ziqiang Wang}
\address{Department of Physics, Boston College, Chestnut Hill, MA 02167}
\date{January 1, 1998}
\maketitle

\begin{abstract}
The transport properties on the two-dimensional surface of 
coupled multilayer heterostructures are studied
in the integer quantum Hall states. 
We emphasize the criticality of the surface state and the
phase coherent transport properties in the thermodynamic limit.
A new, stable numerical algorithm for large scale conductance calculations
in the transfer matrix approach is discussed in detail. 
It is then applied to a directed network model describing the 
quantum mechanical tunneling and impurity scattering of the 
multilayer edge states. We calculate
the two-probe conductance in the direction parallel to
the external magnetic field, its fluctuations and statistical
distributions as a function of the interlayer tunneling strength.
Using finite size scaling, the asymptotic scaling functions of the 
ensemble averaged conductance and the conductance fluctuations
are calculated for a fixed aspect ratio 
and found to be in remarkable agreement with
the analytical results obtained using the supersymmetric
nonlinear $\sigma$-model approach. The conductance distribution
is determined in the quasi-one-dimensional metallic, insulating,
as well as the crossover regime where comparisons are made to that at the
single-layer quantum Hall transition.
We present, for the first time, a detailed study of 
the level statistics in the eigenvalue spectrum of
the transfer matrix. Coexistence of metallic and insulating statistics 
is observed in the crossover regime, which is attributed to the emergence
of a finite range level repulsion in the crossover regime, 
separating the metallic (Wigner-surmise) behavior at small level spacings 
from the insulating (uncorrelated or Poisson) behavior at large level 
spacings. 
\end{abstract}
]

\section{Introduction}
When a two-dimensional electron gas (2DEG) is placed under a strong
perpendicular magnetic field, it exhibits a remarkable 
set of low-temperature magnetotransport behaviors known as the quantum Hall 
effect (QHE) \cite{von,tsui,books}. The main part of the phenomenology 
can be summarized by the existence of (1) new stable phases of matter, 
{\it i.e.} the quantum Hall states, with vanishing dissipation and 
quantized Hall conductances;
(2) continuous, zero temperature phase 
transitions between the quantum Hall states; and
(3) extended, current-carrying, edge states
in the quantum Hall liquids.
These edge states are intrinsically chiral due to the complete breaking of
time reversal symmetry by the external magnetic field.

Two-dimensionality has been regarded as a necessary
condition for observing the QHE. Recently, the question
concerning what happens to the physics, {\it e.g.} the three aspects
mentioned above, associated with the QHE in materials
having engineering dimensions between two and three has received
substantial theoretical and experimental interests.
Integer quantized Hall plateaus have been observed experimentally
in weakly coupled 30 \cite{stormer} and 200 \cite{brooks}
multilayer GaAs/AlGaAs graded structures. In the presence of
finite interlayer tunnelings, the latter are the natural generalizations 
of the QHE above 2D. There are several
important issues \cite{chalkerd,bfisher,wang}: 
(1) the quantization condition
for the Hall resistance; (2) the stable phases of matter and the
nature of the phase transitions; (3) the new physics associated with the
edge states phenomena in the quantum Hall states.

\begin{figure} 
\center 
\centerline{\epsfysize=2.2in 
\epsfbox{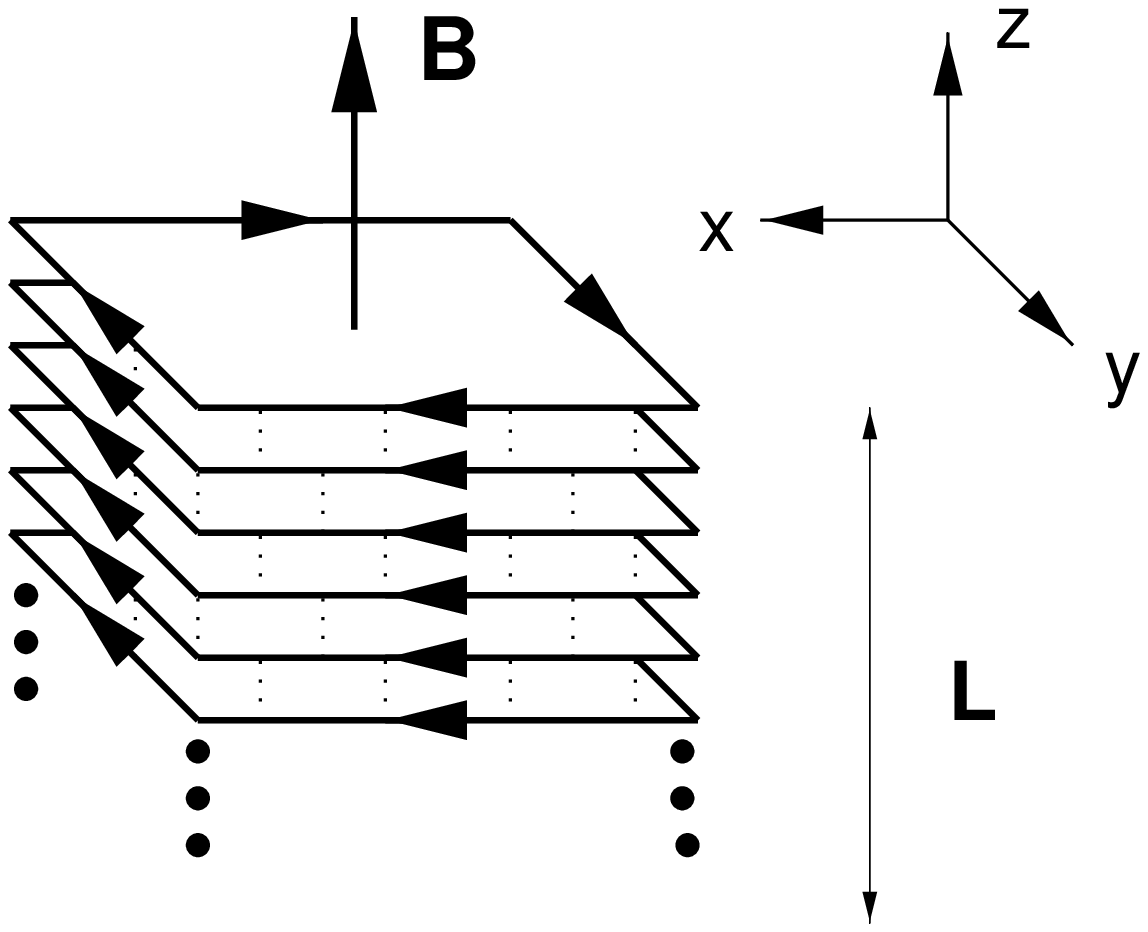}
} 
\begin{minipage}[t]{8.1cm}   
\caption{Schematics of a multilayer quantum Hall sample. The dotted lines
indicate the coupling between the edge states on the surface
due to quantum tunneling.
} 
\label{multilayer}
\end{minipage}   
\end{figure}

In this paper, we shall focus on (3), under the condition that the
multilayer graded heterostructure is in the bulk quantum Hall state.
In this case, the bulk localization length $\xi_{3D}$ is finite and
very short. The edge state in each layer thus decouples from 
the bulk. New physics arises because the finite interlayer tunnelings
on the sample boundaries couple the edge states supported by individual layers
together to form an interesting 2D system on the surface of the
multilayer \cite{chalkerd,bfisher}. This is shown schematically in 
Fig.~\ref{multilayer}.
At this stage, one assumes that the electron-electron interactions can be
ignored and the edge states can be treated as independent electrons
undergoing impurity scattering and interlayer tunneling as described
in Fig.~\ref{multilayer}.

The three most noteworthy novelties of the surface state under such
settings are as follows. (i) It is a 2D chiral system because of the 
unidirectional edge states. (ii) As such, the backscattering
mechanism of 2D weak localization is effectively suppressed.
The surface electrons are in a critical state, in the sense that
the localization length $\xi$ is linearly proportional to the 
linear dimension of the surface \cite{chalkerd}.
(iii) The transport is intrinsically anisotropic. 
Let us consider such a sample surface as shown in Fig.~\ref{multilayer}, with
dimensions $L$ in the interlayer direction and a layer circumference $C$.
In the presence of disorder,
the chiral nature of the edge states keeps the in-plane transport
ballistic with a velocity $v_B$. 
The typical time required for an electron to circumnavigate the
sample is thus given by a ballistic time $\tau_B=C/v_B$.
The transport along the direction parallel to
the magnetic field ($z$-direction) becomes, on the other hand, diffusive
characterized by a diffusion constant $D$ which is controlled by
the strength of the interlayer tunneling $t$.
This leads to a diffusion time $\tau_D=L^2/D$ across the sample 
in the $z$-direction.

\begin{figure}  
\center  
\centerline{\epsfysize=2.4in  
\epsfbox{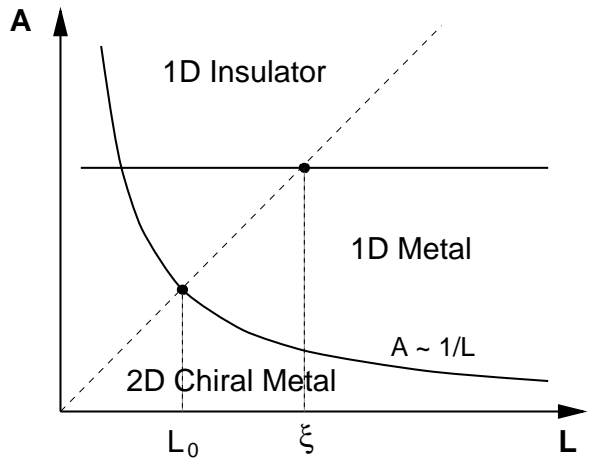}
}
\begin{minipage}[t]{8.1cm}   
\caption{Schematics of the crossover between different phase coherent
transport regimes on the surface (see text). The dashed line corresponds
to a constant circumference $C$, its intersections with the solid crossover
lines determine $L_0$ and $\xi$ respectively.
} 
\label{crossover} 
\end{minipage}   
\end{figure}  

The transport property in the magnetic field direction ($z$) 
has been a recent subject of considerable theoretical 
\cite{chalkerd,bfisher,bfz,kim,mathur,yu,grs1,grs}
and experimental interests \cite{druist}.
The most striking feature is the emergence of several regimes of phase
coherent transport in samples where the phase coherent
length $L_\phi\gg L,C$. These regions are most clearly 
exhibited in the plane spanned by $L$ and the surface aspect ratio,
\eq
A\equiv{L\over C},
\label{aspect}
\ee
as shown schematically in Fig.~\ref{crossover}.
(1) The localization length of the critical state,
$\xi\sim C$, separates the metallic regime for $L<\xi$ from the
quasi-1D insulating regime for $L\gg\xi$.
This crossover is shown by the horizontal line at $A={\rm const.}$
in Fig.~\ref{crossover}, where $L=\xi$.
(2) Next, equating the ballistic time $\tau_B$ and the diffusion time $\tau_D$
leads to a new length scale $L_0\sim \sqrt{C}$. Associated with
the latter is another crossover between a chiral 2D and the 
quasi-1D metallic regime, as shown in Fig.~\ref{crossover} by the curved line
with $A\propto1/L$ on which $L=L_0$.
In the 2D metal regime,
the electrons typically diffuse across $L$ before ballisticly
traverse across the circumference $C$.
(3) Finally, the regime of small $L$ and large $A$ is expected 
to be dominated by ballistic transport.

For fixed dimensions $L$ and $C$, and thus fixed aspect ratio $A$,
the conductance in the $z$-direction
is expected to show smooth crossovers between the
three regimes as a function of the interlayer tunneling $t$.
Alternatively, for a fixed $t$, the crossovers take place
with varying sample geometry. The dashed line in Fig.~\ref{crossover}
shows an example of the latter, along which $C$ is kept fixed.
As $L$ increases and sweeps through $L_0$ and $\xi$, 
transport properties cross over from those of a 2D chiral 
metal to those of the quasi-1D metal and finally to those of
a quasi-1D insulator. 

In this work, we are interested in the transport properties
along the field direction {\it in the thermodynamic limit}, {\it i.e.} when
$C\to\infty$ and $L\to\infty$.
A large scale numerical transfer matrix algorithm will be discussed
and applied to calculate the two-terminal conductance and
its mesoscopic fluctuations in phase coherent samples.
The behaviors in the thermodynamic limit will be obtained
via finite size scaling.
Note that the transport behaviors of the surface state
in the thermodynamic limit depend crucially on how the latter is taken.
In particular, since $\xi\propto C$, the thermodynamic state reached 
along a path with a finite but fixed aspect ratio $A={\rm const.}$ 
corresponds, for any interlayer tunnelings, to the 1D metallic or 
insulating regimes, where $L/L_0\sim L/\sqrt{C}\to\infty$, 
as shown in Fig.~\ref{crossover}.
The 2D chiral metallic regime, on the other hand, can only be reached in the
thermodynamic limit with a vanishing aspect ratio (see Fig.~\ref{crossover})
$A\sim L^{-\alpha}$ and $\alpha\ge1$. 
The scaling forms for the ensemble averaged conductance and its variance
have been written down by Gruzberg, Read, and Sachdev (GRS) \cite{grs1,grs},
\eq
<g>={e^2\over h}\Gamma\left({L\over\xi},{L\over L_0}\right),
\label{gscaling}
\ee
and similarly, the $2n$-th central moment follows,
\eq
<\delta g^{2n}>=\left(e^2\over h\right)^{2n}\Gamma_{2n}
\left({L\over\xi},{L\over L_0}\right).
\label{dgscaling}
\ee
Here, the crossover lengths $L_0$ and $\xi$ depend on $C$ and the
strength of the interlayer tunneling $t$.
Note that the scaling variable $L/\xi$, with $\xi$ being the localization 
length in a finite system, contains the aspect ratio ($A$) 
dependence in the scaling limit where $\xi\propto C$ \cite{thanks}.

In the quasi-1D regime, it was pointed out by GRS that the scaling functions
$\Gamma$ and $\Gamma_2$ should
follow those in a quasi-1D wire \cite{grs1,grs}. The asymptotic
forms of the latter have been obtained exactly by Mirlin {\it et al.} 
using the supersymmetric nonlinear $\sigma$-model \cite{mirlin}.
The exact scaling
functions in the chiral 2D regime are not known. Recently, they
have been studied perturbatively,
and the leading order corrections to the scaling function
have been obtained \cite{grs}. The first comparison of numerical
calculations with these analytical results was made by Cho, Balents, and 
Fisher \cite{chofisher}.
In addition to study the behaviors of the mean conductance and its variance,
we will also analyze the behaviors of the
conductance distribution function and
the level statistics in different regimes and near the crossover.

The rest of the paper is organized as follows. In section II, we
will describe a directed network model for transport
on the surface. We will discuss the properties of the
transfer matrix for chiral transport and compare it to the one
used in the 2D quantum Hall transport. In section III, we will discuss a
new numerical algorithm for large scale conductance calculations.
The method for studying the level
statistics is also explained. We will then present our numerical
results in section IV and conclude in section V together with
a discussion of the open issues.

\section{Directed Network Model}

\begin{figure}   
\center   
\centerline{\epsfysize=2.4in   
\epsfbox{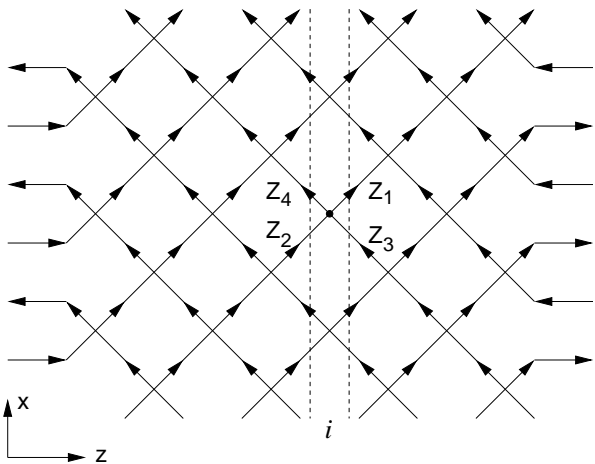}
}
\begin{minipage}[t]{8.1cm}   
\caption{The directed network model.
The links with directed arrows indicate
the propagation of the edge state in individual layers,
whereas the interlayer tunneling takes place at the
nodes. The magnetic field is along the $z$-direction.
} 
\label{network}    
\end{minipage}   
\end{figure}

To facilitate a calculation of the transport properties in the presence
of a smooth-varying impurity potential, we model the motion of the electrons 
on the surface (see Fig.~\ref{multilayer})
by the directed network (DN) model shown in Fig.~\ref{network}
introduced by Chalker and Dohmen \cite{chalkerd,saul}.
The propagation of the chiral edge states is represented by the
directed links in the DN. The impurity scattering is accounted for by
letting the wavefunctions accumulate random Aharonov-Bohm
phases along the links \cite{cc}. The interlayer tunneling between
the edge states take place at the nodes of the DN.

We define the dimensions of the DN in Fig.~\ref{network} as follows:
$L$ along the horizontal ($z$) direction, corresponding to
$L$ columns of nodes where tunneling between adjacent layers
takes place among the $L$ coupled layers; $C$ in the transverse ($x$) 
direction, which amounts to a layer circumference of $C/2$ rows of
nodes. For simplicity, the lattice constants of the DN
will be set to unity such that $L$ and $C$ are dimensionless numbers.
Periodic boundary conditions are applied
in the transverse direction,  whereas two semi-infinite ideal leads
will be attached to the ends for the calculation of 
the two-terminal conductance in the $z$-direction.

The difference of the DN from the Chalker-Coddington network model
for the integer quantum Hall transport \cite{cc} is that the probability 
current on all the links is unidirectional.
As a result, the structure of the transfer matrix describing
a quantum tunneling event at a node is different.
As shown in Fig.~\ref{network}, at each node, there are two incoming
and two outgoing modes. Let us denote the associated probability 
amplitudes by $Z_{\rm L,in}=Z_2$, $Z_{\rm L,out}=Z_4$, 
$Z_{\rm R,in}=Z_3$, and $Z_{\rm R,out}=Z_1$. 
The transfer matrix must be defined in a way such
that the transfer process is multiplicative throughout the DN,
\eq
\pmatrix{Z_{\rm L,out}\cr Z_{\rm L,in}\cr}=
T
\pmatrix{Z_{\rm R,out}\cr Z_{\rm R,in}\cr},
\label{matrixz}
\ee
where $T$ is a $2\times2$ transfer matrix. The current conservation at each
node implies that $T$ must leave $\sum_{i=L,R}\vert Z_{\rm i,in}\vert^2$ 
invariant, {\it i.e.},
$$
\vert Z_{\rm L,in}\vert^2-\vert Z_{\rm L,out}\vert^2 = \vert Z_{\rm R,out}
\vert^2-\vert Z_{\rm R,in}\vert^2.
$$
Thus, $T$ has noncompact unitary symmetry, $T\in U(1,1)$.
With a choice of gauge, one has,
\eq
T=\pmatrix{\sinh\theta & \cosh\theta\cr
\cosh\theta & \sinh\theta\cr},
\label{tchiral}
\ee
controlled by a real tunneling parameter $\theta$. Notice that
$\det T=-1$, {\it i.e.} for the chiral surface transport,
$T$ belongs to the noncompact unitary group with determinant 
negative unity.
In general, $\theta$ is determined by the Fermi energy and the local 
potential. We will take $\theta$ to be the same on all the nodes, corresponding
to uniform interlayer tunnelings. 
We found that a randomly distributed $\theta$ about
a finite mean value across the DN is irrelevant for the asymptotic 
transport properties of the {\it critical} surface states. This situation
is reminiscent of the one encountered in the ordinary network model 
description of the integer quantum Hall transition \cite{cc}. In
the latter case, the randomness in the tunneling amplitude at the saddle 
\twocolumn[
\hsize\textwidth\columnwidth\hsize\csname@twocolumnfalse\endcsname
points of the random potential was found to be irrelevant \cite{lwk,wjl}.

The scattering $S$-matrix, on the other hand, transforms the incoming
amplitudes to the outgoing ones at a node,
\eq
\pmatrix{Z_{\rm L,out}\cr Z_{\rm R,out}\cr}=
\pmatrix{r & t\cr t & -r\cr}
\pmatrix{Z_{L,in}\cr Z_{\rm R,in}\cr},
\label{matrixzs}
\ee
where $t$ and $r$ are the transmission and the reflection
coefficients respectively. 
From Eqs~(\ref{matrixz}) and (\ref{tchiral}), one finds,
\eq
t={1\over\cosh\theta},\qquad r=\tanh\theta.
\label{tran}
\ee
In the rest of the paper, we will study the transport properties
as a function of the single-node interlayer tunneling strength
$t^2$.

It is instructive to compare the transfer matrix in Eq.~(\ref{tchiral})
to the one describing the quantum Hall transport within the layer \cite{cc},
\eq
T_{\rm QH}=\pmatrix{\cosh\theta & \sinh\theta\cr
\sinh\theta & \cosh\theta\cr}.
\label{tqh}
\ee
The transfer matrix in this case belongs to the noncompact unitary
group $U(1,1)$ with determinant $\det{T_{\rm QH}}=1$.
Thus, the two problems are very different and belong to different 
ensembles of random matrices,
except in the pathological limit of $\theta\to\infty$, {\it i.e.}
$r\to1$ and $t\to0$, where $T=T_{\rm QH}$.

Using Eq.~(\ref{tchiral}) as the building block, one can construct
the $C\times C$ transfer matrix for the $i$-th column of the
DN in Fig.~\ref{network},
\eq
T_i=\pmatrix{\pmatrix{e^{i\phi_1} & 0\cr 0& e^{i\phi_2}\cr}T
& & 0\cr
&\ddots& \cr
0& & \pmatrix{e^{i\phi_{C-1}} 
& 0\cr 0& e^{i\phi_{C}}\cr}T \cr}.
\label{tslice}
\ee
Here $\phi_i\in [0,2\pi)$ are the random U(1) phases on the links
of the DN. The transfer matrix for the $i\pm1$-th column has
a slightly different structure,
\eq
T_{i\pm1}=\pmatrix{e^{i\phi_1^\prime}\sinh\theta & & & & 
e^{i\phi_1^\prime}\cosh\theta \cr
& \pmatrix{e^{i\phi_2^\prime} & 0\cr 0& e^{i\phi_3^\prime}\cr}T
& & 0 & \cr
& & \ddots & & \cr
& 0& & \pmatrix{e^{i\phi_{C-2}^\prime}  
& 0\cr 0& e^{i\phi_{C-1}^\prime}\cr}T & \cr
e^{i\phi_{C}^\prime}\cosh\theta & & & &  
e^{i\phi_{C}^\prime}\sinh\theta \cr},
\label{tslice2}
\ee
for periodic boundary conditions. Here $\phi_i^\prime\in [0,2\pi)$ 
are again the random U(1) phases.
\\
]

The total transfer matrix for the entire DN is given by the matrix product,
\eq
\bt=\prod_{i=1}^{L} T_i.
\label{bt}
\ee
Correspondingly, the scattering $S$-matrix has the form
\eq
\bs=\pmatrix{\br & \btt \cr
\btt & -\br \cr},
\label{bs}
\ee
where $\btt$ and $\br$ are the 
$(C/2)\times (C/2)$ transmission and reflection matrices respectively.

Now we discuss the symmetry properties of the transfer matrix product
in Eq.~(\ref{bt}). The current conservation
across the entire DN implies that $\bt$ must leave
$$
\sum_{l=1}^{C/2}\left(\vert Z_{\rm L,in}^{\ l}\vert^2-
\vert Z_{\rm L,out}^{\ l}\vert^2\right)
=\sum_{l=1}^{C/2}\left(\vert Z_{\rm R,out}^{\ l}\vert^2-\vert 
Z_{\rm R,in}^{\ l}\vert^2\right).
$$
Thus $\bt$ forms a noncompact unitary
group $U(C/2,C/2)$ that has the following metric preserving properties:
\eqa
\bt^\dagger \Sigma \bt&=&-\Sigma,
\label{u1}\\
\bt \Sigma \bt^\dagger&=&-\Sigma,
\label{u2}
\eea
where the $C\times C$ matrix $\Sigma$ can be written as a direct
product ($\otimes$),
$$
\Sigma=\tau_z\otimes\bunity,
$$
of the $2\times2$ Pauli matrix $\tau_z$ and the $(C/2)\times (C/2)$
identity matrix $\bunity$.
Notice the minus sign in Eqs~(\ref{u1}) and (\ref{u2}). This is
a consequence of defining the transfer matrix 
that is multiplicative across the DN, as in the single-node case in
Eq.~(\ref{matrixz}). As a result, $\det\bt=-1$, {\it i.e.}
the noncompact unitary group is not volume-preserving.

What we will show next is that the relevant hermitian matrix
$\btdg\bt$ entering the conductance formula (see section III)
has the desired properties.
Substituting Eq.~(\ref{u2}) to the left hand side of Eq.~(\ref{u1}) for
$\Sigma$ leads to,
\eq
(\btdg\bt)\Sigma(\btdg\bt)=\Sigma.
\label{ubt}
\ee
Thus, $\btdg\bt$ forms the volume-preserving U(C,C) group.
Denoting the eigenvalues of the latter by $\{\lambda_i\}$, 
$i=1,\dots,C$, in descending order, $\lambda_i>\lambda_{i+1}$, we
next show that the eigenvalues form reciprocal pairs,
\eq
\lambda_i\cdot\lambda_{C-i+1}=1.
\label{lambda}
\ee
Under a unitary transformation,
$$
\bp=\bu^\dagger (\btdg\bt) \bu, \quad \bu^\dagger\bu=\bunity,
$$
the matrix $\bp$ has the same set of eigenvalues $\{\lambda_i\}$.
Following Ref.\cite{muttalib}, Eq.~(\ref{ubt}) implies that
\eq
\bp^\dagger \cj \bp=\cj,\quad
\cj=\bu^\dagger \Sigma\bu.
\label{p}
\ee
With the following choice of $\bu$,
$$
\bu={1\over\sqrt{2}}\left(\tau_0\otimes\bunity
-i\tau_x\otimes\bunity\right),
$$
the metric preserved by $\bp$ in Eq.~(\ref{p}) becomes
$$
\cj=i\tau_y\otimes\bunity
=\left( \begin{array}{cccccc}
  & 1 & & & & \\
-1 &  & & & & \\
 & &  & 1  & & \\
 & & -1 &  & & \\
 & &  & & \ddots & \\
 & & & & & \ddots
\end{array} \right).
$$
Thus the $\bp$ matrices belong to the symplectic group Sp(C). As a consequence,
the eigenvalues of $\bp$, and thus those of $\btdg\bt$,
form reciprocal pairs as in Eq.~(\ref{lambda}). This
property will be used in the conductance calculations in the
following section.

\section{Numerical Algorithm}

To determine the two-terminal conductance in the
magnetic field direction ($z$), two semi-infinite ideal 
leads are attached at the left and right ends of the disordered
DN in Fig.~\ref{network}. The conductance is then given
by the Landauer formula \cite{fisherlee,bstone},
\eq
g={e^2\over h}\tr[\btt^\dagger \btt],
\label{g}
\ee
where $\btt$ is the $(C/2)\times (C/2)$ transmission matrix defined in 
Eq.~(\ref{bs}). 
Using the unitarity of $\bs$
and the symplectic property of $\btdg\bt$, it is straightforward to show
that Eq.~(\ref{g}) can be written in terms of the transfer
matrices \cite{pichard},
\eq
g={e^2\over h}
\tr\left[{2\over\btdg\bt+\left(\btdg\bt\right)^{-1}+2\cdot\bunity}\right].
\label{g2}
\ee
Further simplication can be made by writing the ordered eigenvalues
($\{\lambda_i\}$) of $\btdg\bt$ as,
\eq
\lambda_{i}=\left\{ \begin{array}{ll}
e^{2\gamma_i}, & i=1,\dots,C/2, \\
e^{-2\gamma_i}, & i=C/2+1,\dots,C. \end{array} \right.
\label{lambdai}
\ee
Evaluating the trace in Eq.~(\ref{g2}) leads to,
\eq
g={e^2\over h}\sum_{i=1}^{C/2}{1\over\cosh^2(\gamma_i)}.
\label{g3}
\ee
In the remainder of the paper, we will set the conductance unit 
$e^2/h\equiv1$. 

The two-terminal conductance can be calculated
from Eq.~(\ref{g3}) or Eq.~(\ref{g}) for a given sample
with a disorder realization. The conductance distribution is then
obtained by repeating the calculation for a large ensemble of
samples with microscopically different disorder realizations.
The ensemble averaged conductance, conductance fluctuations,
and in fact all statistical properties of the conductance can be 
obtained from the conductance distribution.
Nevertheless, this conceptually straightforward procedure has been hindered
by the numerical difficulties involved in evaluating the product of
a large number of random transfer matrices (see {\it e.g.} Eq.~(\ref{bt})).
Unlike the localization length, which can be obtained in the
quasi-1D limit of a long stripe or a cylinder using Oseledec's theorem
\cite{os}, the conductance of a $C\times L$ sample requires the
knowledge of the product matrix of $2L$ random transfer matrices of
dimension $C\times C$ (Eq.~(\ref{g2})), or equivalently, that of
all its eigenvalues (Eq.~(\ref{g3})). For large $L$ and $C$, 
direct calculation of the product is impossible as one quickly
looses numerical stability in multiplying the transfer matrices due to the
inevitable dynamical range problem. The latter refers to the inability of
today's computers to keep within accuracy of exponentially growing
and exponentially decreasing matrix elements. 
As a result, the applicability of this
method has been limited to small system sizes.

Now we discuss a stable numerical algorithm  that removes the
above difficulty. We shall first prove a theorem, which is then
applied to the evaluation of the eigenvalues $\gamma_i$
in Eq.~(\ref{g3}).

Consider a general hermitian matrix $H$ with non-degenerate 
eigenvalues.

\bigskip

\noindent{\bf Theorem 1 }
\begin{it}
The $n$-th power of $H$, {\it i.e.} the matrix $H^n$ can be written 
in the form:
\eq
H^n=U_n D_n R_n,
\label{hn}
\ee
where, in the limit of large $n$, (i) $U_n$ is a unitary matrix of which
the columns converge to the eigenvectors of $H$; (ii) $D_n$
is a diagonal matrix 
and the eigenvalues of $H$ is given by $D_n D_{n-1}^{-1}$
in descending order; (iii) $R_n$ is
a right triangular matrix with unity on the diagonals that converges
to a limiting matrix of that structure.
\end{it}

\noindent{\it Proof:} Let us begin with the fact that the hermitian
matrix $H$ can be factorized according to,
\eq
H=U_1D_1R_1.
\label{h1}
\ee
This UDR-decomposition is identical to the well-known Gram-Schmidt
procedure for orthonormalizing the column vectors in $H$.
Apply $H$ to Eq.~(\ref{h1}), one has,
\eq
H^2=HU_1D_1R_1=U_2 D_2^\prime R_2^\prime D_1R_1.
\label{h2}
\ee
Here we have done another UDR-decomposition to the product
$HU_1$. Thus $R_2^\prime$ is right triangular with unit
diagonal elements. Now we interchange $R_2^\prime$ and $D_1$,
\eq
R_2^\prime D_1=D_1 R_2^{\prime\prime},
\ee
where $R_2^{\prime\prime}$ is right-triangular with non-vanishing elements,
\eq
R_2^{\prime\prime}(i\le j)={D_1(j,j)\over D_1{(i,i)}}R_2^\prime(i,j).
\ee
Thus Eq.~(\ref{h2}) becomes,
\eq
H^2=U_2D_2R_2,
\label{h22}
\ee
where 
\eq
D_2=D_2^\prime D_1,\quad R_2=R_2^{\prime\prime}R_1,
\ee
are diagonal and right-triangular matrices respectively.
Repeating the steps from Eq.~(\ref{h1}) to Eq.~(\ref{h22}) leads to
\eq
H^n=U_n D_n R_n, 
\label{hnn}
\ee
and the following recursion relations,
\eqa
D_n&=&D_n^\prime D_{n-1},
\label{recursion1} \\
R_n&=&R_n^{\prime\prime}R_{n-1}.
\label{recursion2}
\eea

Eq.~(\ref{hnn}) is a matrix-factorized form of the product matrix
$H^n$. The unitary matrix $U_n$ is necessarily 
well conditioned. We will show below that the right-triangular matrix
$R$ is well defined and converges to a limiting matrix of that structure
exponentially fast with increasing $n$. The large variations in 
the size of the matrix elements, which limit the dynamical range
of the matrix multiplications, are contained solely in the elements of
the diagonal matrix $D_n$. In fact the Gram-Schmidt orthonormalization
procedure ensures that, in the limit of large $n$, the columns of the
unitary matrix $U_n$ project onto the eigenvectors of $H$.

Next, we demonstrate the convergence of $R_n$ and obtain the eigenvalues
of $H$. The important point is that, in each step, the UDR-decomposition
is done by orthonormalizing the $j$-th column with respect to all columns 
of index $i<j$. As a result, the diagonal elements of $D_n$ are
accumulated in descending order, {\it i.e.} $D_n(i,i)\gg D_n(j,j)$
for $j>i$. Carrying out one more multiplication,
\eq
H^{n+1}=HU_n D_n R_n=U_{n+1}D_{n+1}^\prime R_{n+1}^\prime D_n R_n.
\label{hnp1}
\ee
Interchanging $R_{n+1}^\prime$ with $D_n$,
\eq
R_{n+1}^\prime D_n=D_n R_{n+1}^{\prime\prime},
\ee
where
\eq
R_{n+1}^{\prime\prime}(i,j)=R_{n+1}^\prime(i,j){D_n(j,j)\over
D_n(i,i)}.
\label{rddr}
\ee
Since $R(i,j)$ is nonzero for $j\ge i$, Eq.~(\ref{rddr})
shows that the off diagonal elements ($j>i$) are suppressed
exponentially with increasing $n$ by the ratio of $D_n(j,j)/D_n(i,i)$,
while the diagonal elements remain to be unity.
Thus, when $n$ is large, $R_{n+1}^{\prime\prime}$ approaches the 
identity matrix provided that there is no degenerate elements in $D_n$.
The convergence of the $R$ matrix follows since
$R_{n+1}=R_{n+1}^{\prime\prime}R_n\to R_n$ for large $n$.
Going back to Eq.~(\ref{hnp1}), we have
\eq
H^{n+1}=H U_n D_n R_n= U_{n+1} D_{n+1} R_{n+1}.
\label{hnp11}
\ee
Using the fact that $R_{n+1}\to R_n$, Eq.~(\ref{hnp11}) implies,
\eq
U_{n+1}^\dagger H U_n=D_{n+1} D_n^{-1}.
\ee
Thus we have proven that for large $n$, the eigenvalues of
$H$ is given in descending order by the diagonal matrix $D_{n+1}D_n^{-1}$,
and the columns of the unitary matrices $U_n$ and $U_{n+1}$
converge to the corresponding eigenvectors. 
In the absence of nearly degenerate eigenvalues, the procedure 
converges exponentially fast, such that in practice $n$ need not
be too large to obtain the desired accuracy.

It is straightforward to construct an algorithm for 
evaluating the eigenvalues of the transfer matrix product in 
the conductance Eq.~(\ref{g3}) following the
steps of the proof described above. Specifically, we consider $H$ as defined by
\eq
H=\btdg\bt=T_1^\dagger T_2^\dagger\cdots T_L^\dagger T_L\cdots
T_2 T_1.
\label{ts}
\ee
The eigenvalues of $H$ is obtained by the above UDR-decomposition of
\eq
H^{n}=\left(\prod_{i=L}^1 T_i^\dagger \prod_{i=1}^L T_i\right)^n.
\label{hnt}
\ee
In practice, the UDR-decomposition need not be performed after every 
multiplication, as this is the most time-consuming part of the
computation. It scales at best with $C^\alpha$, $2<\alpha<3$.
The number of possible direct multiplications without losing accuracy,
$N_d$, depends on the size of the matrix elements in $T_i$, thus the
tunneling parameter $t$, and weakly on the size of the matrices.
\\

\noindent{\sl Algorithm:}
\begin{enumerate}
\item Carry out $N_d$ steps of direct matrix multiplications 
in Eq.~(\ref{hnt}).
\item Perform a UDR-decomposition as described above. 
\item Repeat the above procedure for a total of $(n-1)\times(2L)$ 
multiplications for the convergence of the factorized matrices.
\item For the last $2L$ multiplications, store and accumulate
the logarithm of the elements in the diagonal matrix for each of the 
$N_{udr}=2L/N_d$ UDR-decompositions as
\eq
2\gamma_i=\log\lambda_i=
\sum_{l=1}^{N_{\rm udr}}\log D_l(i,i).
\label{gamma}
\ee
For large enough $n$, $\{\lambda_i\}$ gives the set of eigenvalues
for $H$ in descending order.
\item To check convergence, apply $H$ again to the unitary matrix
at the end of the last UDR-decomposition. Repeating step 4, one has
\eq 
2\gamma_i^\prime=\log\lambda_i^\prime= 
\sum_{l=1}^{N_{\rm udr}}\log D_l(i,i). 
\label{gammap} 
\ee
The difference between $\{\gamma_i\}$ and $\{\gamma_i^\prime\}$
can be compared to the convergence criterion, and thus
determine the appropriate power $n$.
\item The conductance can be readily calculated from Eq.~(\ref{g3}).
\end{enumerate}

We emphasize that the applicability of this stable algorithm for conductance 
calculations is very general. It is particularly useful for 
studying phase-coherent transport properties in disordered macroscopic and 
mesoscopic systems in the thermodynamic limit. 
The only limitation in this case is the CPU time.
It has been successfully applied to study the conductance
and conductance fluctuations in the 2D integer QHE \cite{wjl,jw}.
The specific choice of the two parameters, $n$ and $N_d$, depends
on the system under investigation (see section IV). 
In general, it is more convenient to choose $N_d$ such that 
$N_{udr}$ is an integer. Otherwise, an additional UDR decomposition
must be performed at the end of each set of the $2L$ multiplications.
As evidenced in the procedure leading to the proof of the above algorithm, 
essentially, only $U$ and $D$ need to be kept, and $R$ can be discarded 
in each of the UDR decomposition. Thus, even the standard Gram-Schmidt
procedure works for the present purposes.
In the next section, we will apply this
technique to study the statistical properties of the transport on
the surface of a layered quantum Hall state.

\section{The Quasi-1D Regime}

To study the quantum interference effects in 
the conductance, it is convenient to introduce the Ohmic conductance
in the classical limit, $g_{\rm ohm}$. The latter can be 
obtained analytically for the DN in Fig.{\ref{network}} by summing over
all possible non-crossing Feynman paths
contributing to the $z$-axis transport, while ignoring
the nontrivial interference effects associated with the intersecting
ones winding at least once around the circumference of the DN
\cite{chalkerd,chofisher}. This leads to the averaged conductance,
\eq
g_0={C\over2L}{t^2\over 1-t^2(1-1/L)},
\label{g0}
\ee
where $t^2$ is the interlayer tunneling amplitude defined in Eq.~(\ref{tran}).
The $L$-dependence in the denominator in Eq.~(\ref{g0}) is indicative of 
the ballistic contribution that dominates in samples with
$L\ll t^2/(1-t^2)$. Thus, the crossover from ballistic to diffusive
transport is rather slow if $t^2$ is close to one.

In the limit $L\to\infty$, one recovers the Ohmic behavior in the
$z$-direction from Eq.~(\ref{g0}),
\eq
\gohm={C\over L}\sigma, \quad \sigma={1\over2}{t^2\over 1-t^2}.
\label{gohm}
\ee
As a result, the dependence of measurable quantities on the microscopic 
parameter $t^2$ can be favorably replaced by the dependence on
the Ohmic conductance $\gohm$ or the conductivity $\sigma$.
For example, as discussed in Refs~\cite{bfz,grs1,grs}, 
the 1D to 2D crossover length $L_0$ in the scaling
equations (\ref{gscaling}) and (\ref{dgscaling}) can be expressed as
\eq
L_0=\sqrt{\sigma C},
\label{l0}
\ee
and the localization length $\xi$ as
\eq
\xi=2\sigma C.
\label{xi}
\ee

We consider square DNs with aspect ratio $A=1$,
unless otherwise specified. The asymptotic behavior of the transport 
properties in the thermodynamic limit 
taken with $L=C\to\infty$ will be determined 
using the finite size scaling (FSS) analysis.
In this case, 
\eq
{L\over L_0}=A\sqrt{{C\over\sigma}}\to\infty,
\label{loverl0}
\ee
such that the system is always in the quasi-1D regime in the thermodynamic
limit (see Fig.~\ref{crossover}). In fact, this is true so long as the 
latter is taken with a fixed finite aspect ratio, $A={\rm const.}$.
Nevertheless,
\eq
{L\over\xi}={A\over 2\sigma},
\label{loverxi}
\ee
which allows one to study the crossover between the quasi-1D metallic
($L/\xi\ll1$) and insulating ($L/\xi\gg1$) regimes as a function of
the interlayer tunneling in the thermodynamic limit.

\subsubsection{Conductances and Conductance Fluctuations}

For a given disorder realization, we calculate the
conductance from the Landauer-formula in the form of Eq.~(\ref{g3})
using the algorithm described in the last section.
Next the conductances of a large ensemble containing $5,000$ to $10,000$ 
microscopically different samples are calculated.
This is done for a sequence of sample sizes of $8\times8$
to $96\times96$ to facilitate the FSS analysis.
The typical parameters used are the following. The direct number of
multiplications $N_d$ ranges from $8$ for large $t^2$ (approaching
one from below) to $64$ for small $t^2$ (approaching zero).
The power $n$ is typically less than $50$ which is more than enough to 
suppress the systematic errors in the obtained eigenvalues below
the statistical errors associated with ensemble averaging.
The latter is typically less than $1\%$ for our data. 

We first present the results for the ensemble averaged conductance.
In the quasi-1D regime, the scaling function in Eq.~(\ref{gscaling})
becomes, in the asymptotic limit,
\eq
<g>=\Gamma\left({L\over\xi},\infty\right).
\label{gscaling1d}
\ee
For fixed aspect ratio, Eq.~(\ref{loverxi}) shows that $L/\xi$ is
only a function of $t^2$. 

\begin{figure}  
\center  
\centerline{\epsfysize=2.8in  
\epsfbox{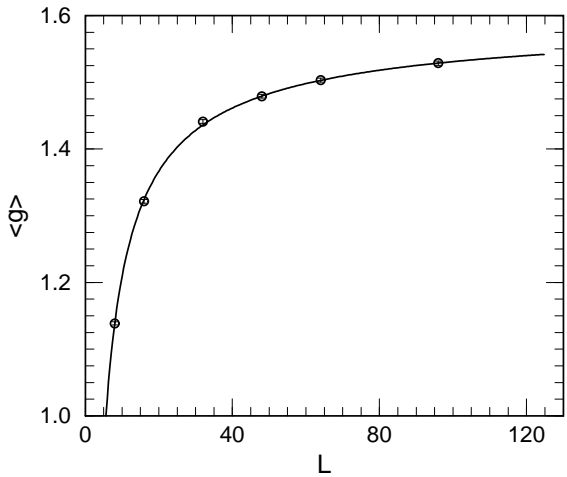}}  
\begin{minipage}[t]{8.1cm}   
\caption{Finite size scaling of the averaged conductance in
the quasi-1D metallic regime at $L/\xi=0.31$.
The solid line is a fit to Eq.~(\ref{fss}), with $y_{\rm irr}
=0.74\pm0.03$ and the asymptotic value $<g>=1.60\pm0.01$.
} 
\label{gvsl}    
\end{minipage}   
\end{figure}

\begin{figure}  
\center  
\centerline{\epsfysize=2.8in  
\epsfbox{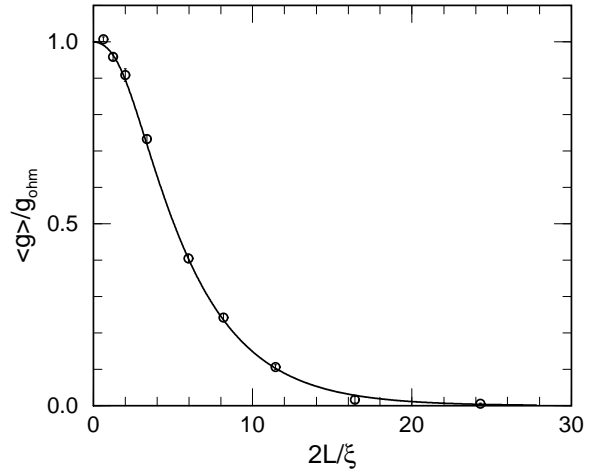}}  
\begin{minipage}[t]{8.1cm}  
\caption{The averaged conductance normalized by $g_{\rm ohm}$
as a function of $2L/\xi$. The data points are the thermodynamic
limit values obtained using finite size scaling.
The solid line is the analytical curve for
a quasi-1D wire obtained by Mirlin {\it et al.}}
\label{gplot}   
\end{minipage}  
\end{figure}

As an example, the calculated
$<g(t,L)>$ is shown in Fig.~\ref{gvsl} as a function of $L$ at
$t^2=0.76$ or $L/\xi =0.31$, which is in the metallic regime.
The apparent size dependence is due to the corrections to scaling.
To extract the asymptotic value, we perform a FSS analysis according to,
\eq
<g(t,L)>=<g(t)>-\zeta_{\rm irr}{1\over L^{y_{\rm irr}}},
\label{fss}
\ee
where $y_{\rm irr}=0.74\pm0.03$ is the dimension of the leading irrelevant 
operator that controls the corrections to scaling and $\zeta_{\rm irr}$ is
a non-universal constant introduced by the conjugate finite length scale.
Following this procedure, the asymptotic values of the conductance are 
obtained for different values of $t^2$. 
In Fig.~\ref{gplot},
the conductance normalized by the Ohmic value, $<g>/\gohm$,
is plotted as a function of $L/\xi$ in accordance with
Eq.~(\ref{gscaling1d}).
The solid line is the analytical scaling function $\Gamma(L/\xi)$
obtained by Mirlin {\it et al.} for a quasi-1D wire using the
supersymmetric non-linear $\sigma$-model \cite{mirlin}.
The agreement throughout the entire regime is remarkable, which 
is in accordance with the analytical work of GSR \cite{grs1}.
It is important to note that such an excellent agreement can only
be obtained for the conductances in the thermodynamic limit
by the FSS analysis of large sample sizes. In fact, the nature of
the finite size corrections is quite different
on the metallic and the insulating side of the crossover.
Let us define
\eq
\beta(t)\equiv {d\over dL} <g(t,L)>.
\label{beta}
\ee
One finds that $\beta(t)\to0$ as $L\to\infty$ consistent with the
fact that the system is critical for all $t^2$ \cite{chalkerd}. 
However, we find that there exists
a well defined tunneling amplitude, $t_{\rm cr}^2\simeq0.37$, such that
$\beta(t)\to0^+$ for $t^2>t_{\rm cr}^2$, whereas 
$\beta(t)\to0^-$ for $t^2<t_{\rm cr}^2$. It is thus natural to
associate the existence of $t_{\rm cr}^2$ with a sharp crossover 
from the metallic
to the insulating regime. The corresponding crossover conductivity
is $\sigma_{\rm cr}\simeq0.3$ and $(L/\xi)_{\rm cr}\simeq1.7$.
These observations provide an explanation of the systematic
deviations from the analytical curve observed numerically in 
relatively small systems \cite{chofisher}.

\begin{figure}   
\center   
\centerline{\epsfysize=2.8in
\epsfbox{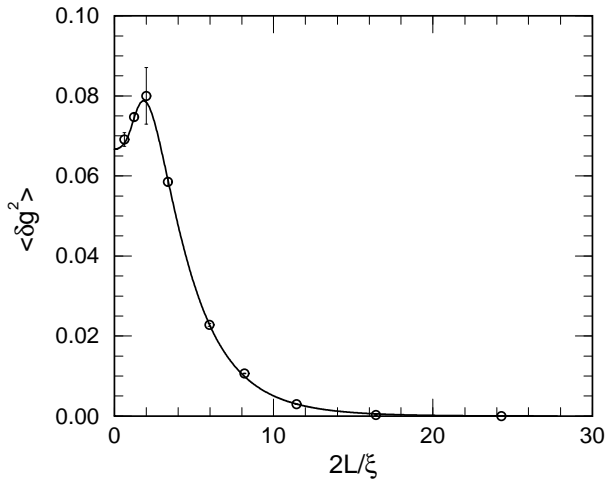}}   
\begin{minipage}[t]{8.1cm} 
\caption{The variance of the conductance in the asymptotic limit
as a function of $2L/\xi$. The solid line is the analytical curve
of $\Gamma_2$ for a quasi-1D wire obtained by Mirlin {\it et al.}.
}
\label{dgplot}  
\end{minipage} 
\end{figure}

Next we present the results for conductance fluctuations about the
ensemble-averaged values.
Once again, we observe significant size dependence in the
variance of the conductances. The latter is indicative of the fact that
the universality of these conductance fluctuations is a property
of the quantum transport at criticality \cite{wjl,jw,grs},
and is outside the context of the conventional universal conductance
fluctuations in diffusive metals \cite{leestone}.
Eq.~(\ref{dgscaling}) has now the asymptotic form,
\eq
<\delta g^2>=\Gamma_2\left({L\over\xi},\infty\right).
\label{dgscaling1d}
\ee
The asymptotic values of the variance
are plotted in Fig.~\ref{dgplot} as a function of $L/\xi$ to
display the scaling function $\Gamma_2$ in Eq.~(\ref{dgscaling1d}).
The solid line is the analytical result of $\Gamma_2$ obtained by Mirlin
{\it et al.} \cite{mirlin}. As in the case of the ensemble-averaged
conductance, the agreement for the asymptotic variance of the conductance
is excellent for the entire quasi-1D regime.
In the limit of small $L/\xi$, the asymptotic variance approaches the
universal value $<\delta g^2>=1/15$, despite the vanishingly 
small fluctuations at small $L$ where the transport is ballistic.
In the opposite limit of large $L/\xi$, the system is in the
1D insulating regime where the absolute fluctuations decay
exponentially. Nevertheless, it is important to note that in the
insulating regime,the averaged conductance itself decays exponentially
\cite{mirlin},
\eq
<g>\propto (\xi/L)^{3/2}e^{-L/2\xi},
\ee
and it is therefore not meaningful to focus on the absolute values
of the variance. In fact, a well known result is that the relative
fluctuations are exponentially large, 
\eq
{\sqrt{<\delta g^2>}\over <g>}\propto e^{L/4\xi},
\ee
deep in the quasi-1D insulating regime.
%
%

We emphasize that this is the first time that conductances
and conductance fluctuations are numerically evaluated in
the thermodynamic limit in this quasi-1D regime. The agreement between the 
numerical data obtained from the DN model and the analytical
results from the supersymmetric nonlinear $\sigma$-model is,
as far as we know, unprecedented.

\subsubsection{Higher Moments and Conductance Distributions}

Next, we present the results for the conductance distributions
in different regimes and provide a statistical description of the distribution
functions, $P(g)$. This will be done by focusing on (1) the largest 
square samples we have studied, {\it i.e.} $96\times96$ DNs, (2) three
values of $L/\xi=0.31$, $1$, and $4.09$ that are
typical representations of the metallic, crossover, and 
insulating regimes. 

\begin{figure}    
\center    
\centerline{\epsfysize=2.8in    
\epsfbox{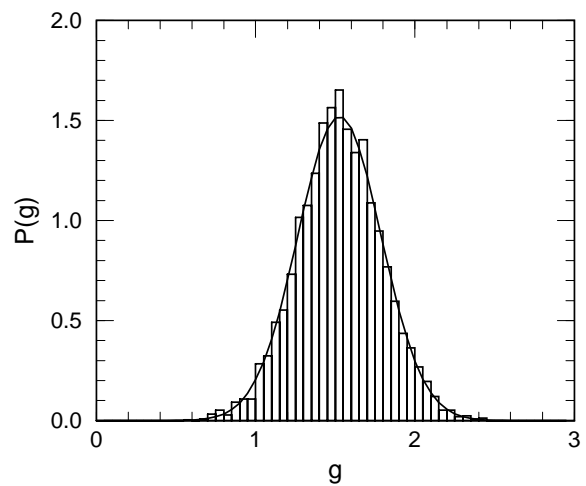}}   
\begin{minipage}[t]{8.1cm}   
\caption{The conductance distribution $P(g)$ 
in the quasi-1D metallic regime at $L/\xi=0.31$. The solid line
is a Gaussian fit to the histogram.
} 
\label{distg1}   
\end{minipage}   
\end{figure}

\begin{figure}     
\center     
\centerline{\epsfysize=2.8in     
\epsfbox{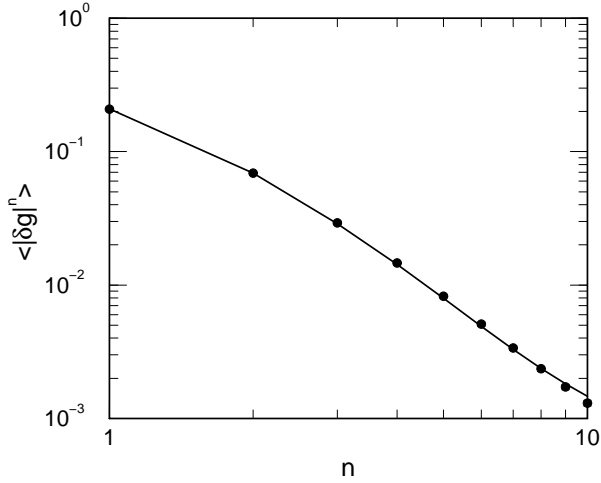}}    
\begin{minipage}[t]{8.1cm}    
\caption{Comparison of the computed central moments of the conductance 
(solid circles) with those determined by the Gaussian fit (solid line)
in Fig.~\ref{distg1}.
}  
\label{distg1mnt}    
\end{minipage}    
\end{figure}

\begin{figure}     
\center     
\centerline{\epsfysize=2.8in     
\epsfbox{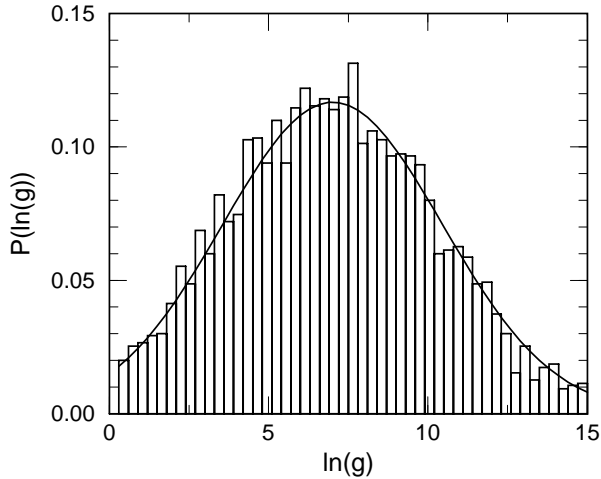}}   
\begin{minipage}[t]{8.1cm}   
\caption{The distribution of the logarithm of the conductance
in the quasi-1D insulating regime at $L/\xi=4.09$. The solid line
is a fit to the log-normal distribution.
} 
\label{distg2}    
\end{minipage}   
\end{figure}

\begin{figure}      
\center      
\centerline{\epsfysize=2.8in      
\epsfbox{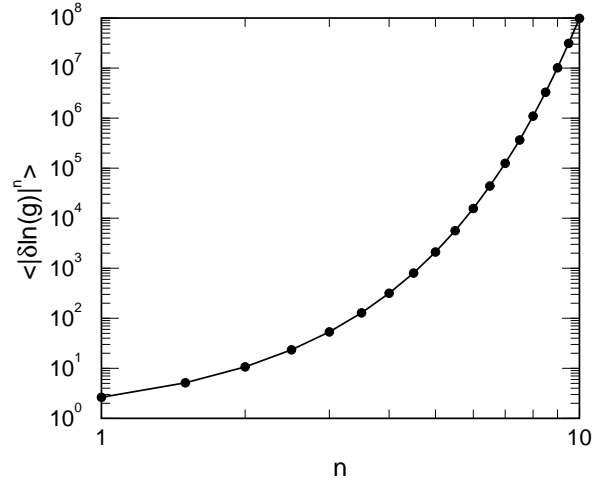}}     
\begin{minipage}[t]{8.1cm}     
\caption{Comparison of the calculated central moments of $\ln g$
(solid circles) with those from the fitted log-normal
distribution (solid line) in Fig.~\ref{distg2}. 
}   
\label{distg2mnt}     
\end{minipage}     
\end{figure}

(i) In the metallic regime, the distribution of the conductance
for $5,000$ samples is shown in Fig.~\ref{distg1}.
The solid line is a fit to the Gaussian distribution 
\eq
P(g)={1\over\sqrt{2\pi\langle\delta g^2\rangle}}e^{-{(g-\langle g\rangle)^2
\over2\langle\delta g^2\rangle}}.
\label{normal}
\ee
The central-moments, $<\vert\delta g\vert^n>$,
are computed for up to $n=10$ from the data and compared to
those determined by the Gaussian fit in Eq.~(\ref{normal}).
The good agreement shows that the conductances are close to
being normal distributed in the metallic regime. The deviation
at large values of $n$ from the Gaussian moments observable
in Fig.~(\ref{distg1mnt}) is indicative of the log-normal tails
in the conductance distribution function \cite{lerner}.
(ii) In the insulating regime, we plot the distribution of the logarithm
of the conductance in Fig.~\ref{distg2} obtained from $5,000$ samples.
Here the solid line is a fit to the log-normal distribution,
\eq
P(\log g)={1\over\sqrt{2\pi\langle\delta \log^2 g\rangle}}
e^{-{(\log g-\langle\log g\rangle)^2
\over2\langle\delta \log^2 g\rangle}}.
\ee
The higher central-moments obtained from our data and the 
fitted log-normal distribution are compared in Fig.~\ref{distg2mnt}.
The excellent agreement confirms that the conductance
follows a log-normal distribution in the quasi-1D insulating regime.
Thus, instead of the conductance, it is the logarithm of the conductance
that is a self-averaging quantity.

\begin{figure} 
\center 
\centerline{\epsfysize=2.8in 
\epsfbox{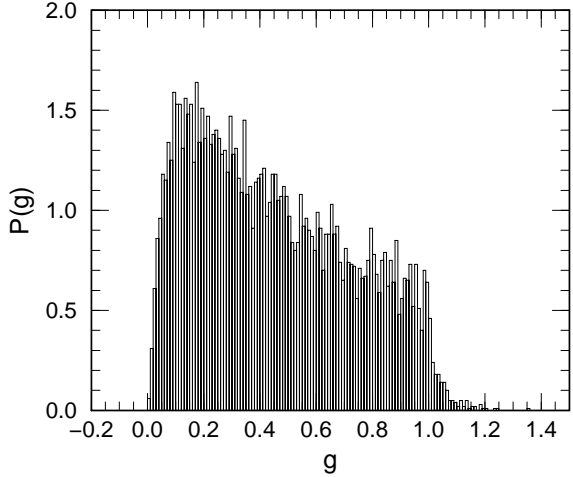}} 
\begin{minipage}[t]{8.1cm}   
\caption{The conductance distribution in the 
crossover regime at $L/\xi=1.0$. Notice that $P(g)$ is broadly
distributed between $0$ and $1$.
} 
\label{distg3}
\end{minipage}   
\end{figure}

\begin{figure}      
\center      
\centerline{\epsfysize=2.8in      
\epsfbox{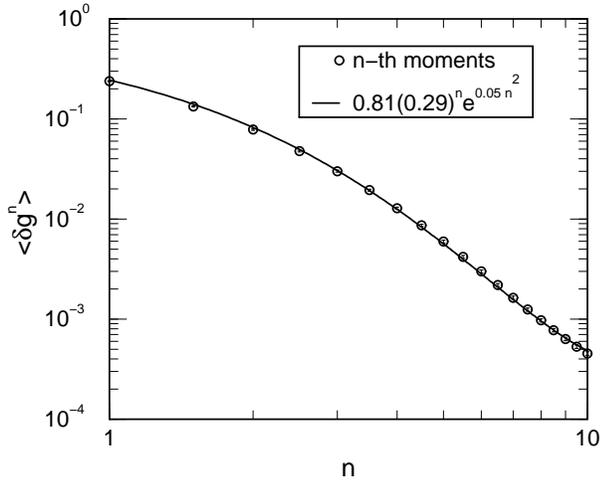}}     
\begin{minipage}[t]{8.1cm}     
\caption{The $n$-th central moments of the conductances in Fig.~\ref{distg3}.
The solid line is a fit to Eq.~(\ref{tail}).
}   
\label{distg3mnt}     
\end{minipage}     
\end{figure}

\begin{figure} 
\center 
\centerline{\epsfysize=2.8in 
\epsfbox{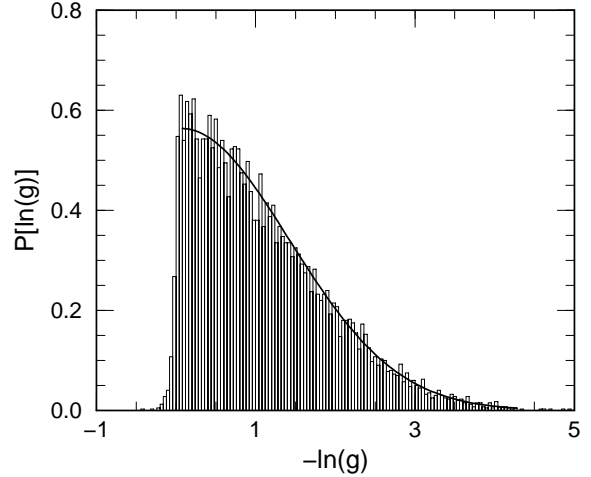}}
\begin{minipage}[t]{8.1cm}   
\caption{The distribution of $\ln(g)$ in the crossover regime 
(see Fig.~\ref{distg3}) 
showing the skewed log-normal distribution (solid line).
} 
\label{distg3log}
\end{minipage}   
\end{figure}

(iii) It is important to ask how the statistical distribution
changes in the crossover regime from the metallic and insulating behaviors.
In this case, we focus on $L/\xi=1$ and $\sigma=0.5$. 
The conductances of a large 
ensemble of $10,000$ $96\times96$ samples were calculated to achieve 
better statistics.
The conductance distribution, $P(g)$, is plotted in Fig.~\ref{distg3}.
Remarkably, the conductance is broadly distributed between
$0$ and $1$, analogous to the behavior of $P(g)$ found
in the Chalker-Coddington network model at the critical point of the 
integer quantum Hall transition \cite{wjl,jw,cho}.
The $n$-th order central-moments are shown in Fig.~\ref{distg3mnt} up
to $n=10$, which can be very well fitted by
\eq
<\delta g^n>= av^ne^{un^2},
\label{tail}
\ee
with $(a, v, u)=(0.81\pm.01,0.29\pm.01,0.050\pm.001)$.
Since Eq.~(\ref{tail}) is indicative of a distribution
with log-normal tails \cite{lerner}, in Fig.~\ref{distg3log},
we plot the distribution of the logarithm of the
conductance, $P[\log(g)]$. 
The latter turns out to show a skewed log-normal distribution as a result
of the sharp fall off of $P(g)$ close to $g=1$.
The conductance can be described
surprisingly well by a log-normal
distribution as shown by the solid line in Fig.~\ref{distg3log}.

We point out the interesting similarity between the 
above features in the crossover regime (iii)
and those found at the 2D integer quantum Hall 
transitions \cite{wjl,jw}. 
A possible scenario to explain this is the following.
The critical properties of the quantum Hall transition are described
by those of the Chalker-Coddington (CC) network model at criticality
\cite{cc,lwk,lwhub}. As pointed out in Refs \cite{lwhub,wjl},
there exists a critical manifold, corresponding to a line
of fixed points, for the CC network model that has the same tunneling
parameter at each node (see Eq.~(\ref{tqh})).
One can show that, as the tunneling parameter is varied along this
line, the critical conductance exhibits a crossover from 2D metallic
to quasi-1D insulating behaviors \cite{vwtobe}. The conventionally studied
quantum Hall transition corresponds to a particular point 
on the critical line selected by requiring an additional
$90$ degrees rotational invariance of the network. The latter
happens to be in the crossover regime
with $L/\xi=\ln(1+\sqrt{2})=0.8814$ \cite{lwk} and $\sigma_c=0.58\pm.03$.
It is possible that the distribution of the conductance
at the critical point in the regime of crossover between
metallic and insulating behaviors possesses the universal features 
that have been observed in both of the systems discussed above.

\subsubsection{Level Statistics}

It is well known that the difference in the statistical properties
of the conductance in disordered metals and insulators originates
from that of the statistical distribution of the eigenstates.
A natural quantity to study is the distribution of the 
normalized (by the averaged value) energy level spacings, often
called the level statistics.
Traditionally, one applies the random matrix theory developed by
Wigner to study the level spacing distribution of an ensemble of
random Hamiltonian matrices \cite{rmtbook}. More recently, it has become clear,
by the Coulomb gas analogy and the maximum entropy hypothesis
\cite{muttalib,pichardstone}, that the same theory describes the
level statistics of the eigenvalues of the transfer matrix. The
latter, in contrast to those of the Hamiltonian matrix, correspond to
the {\it scattering} eigenstates and are directly related
to the Landauer formulation of the conductance
by Eq.~(\ref{g3}).

Here we present our analysis of the transfer matrix level statistic in 
the quasi-1D metallic, insulating, and the crossover regimes of the 
multilayer quantum Hall surface states. Following Eq.~(\ref{lambdai}),
let us introduce the normalized eigenvalues for a given sample,
\eq
\epsilon_i\equiv{\gamma_i\over L}, \quad i=1,\dots, C.
\label{ei}
\ee
The normalized level spacing between adjacent eigenvalues is given by,
\eq
s={\epsilon_{i+1}-\epsilon_i\over\Delta},
\label{s}
\ee
where $\Delta$ is the averaged level spacing,
\eq
\Delta={1\over C-1}<\sum_{i=1}^{C-1} (\epsilon_{i+1}-
\epsilon_i)>.
\label{delta}
\ee

According to the random matrix theory,
the behavior of the level distribution, $P(s)$,
is determined by the level correlations.
In the metallic phase with extended eigenstates, 
the level repulsion is strong and long-ranged, $P(s)$ follows the 
Wigner hypothesis,
\eq
P_{\rm metal}(s)=As^\beta e^{-Bs^\alpha},
\label{ws}
\ee
with $\alpha=2$. In the unitary universality class, $\beta=2$, and
$A={32/\pi^2}$ and $B=4/\pi$ from normalization.
On the other hand, in the insulating regime with exponentially localized 
states, the levels are uncorrelated, the asymptotic behavior
of $P(s)$ for large $s$ follows the Poisson distribution,
\eq
P_{\rm insu} (s)\propto e^{-ks}, \quad s\gg1,
\ee
where $k$ is a positive constant.
However, the behavior of $P(s)$ at the critical point of 
metal-insulator transitions, or in the crossover regime between
metals and insulators is an unresolved issue under
active current debate \cite{shk,kramer,aronov,ono,kawa}.

From the set of the calculated eigenvalues $\{\gamma_i\}$, we now study the
properties of the distribution $P(s)$. 
Notice that in general, the eigenvalue distribution $P(\gamma)$ 
is not uniform.
In order to obtain a meaningful $P(s)$, one has to,
as is normally done \cite{pichard2,ono,kawa}, 
unfold the spectrum of $P(\gamma)$
according to
\eq
\gamma^\prime=\int_{-\infty}^{\gamma^\prime}P(\gamma) d\gamma,
\label{unfolding}
\ee
such that the distribution $P(\gamma^\prime)$ is uniform.
The level spacings are then obtained by replacing $\gamma$ with
$\gamma^\prime$ in Eq.~ (\ref{ei}).

\begin{figure}  
\center  
\centerline{\epsfysize=2.8in  
\epsfbox{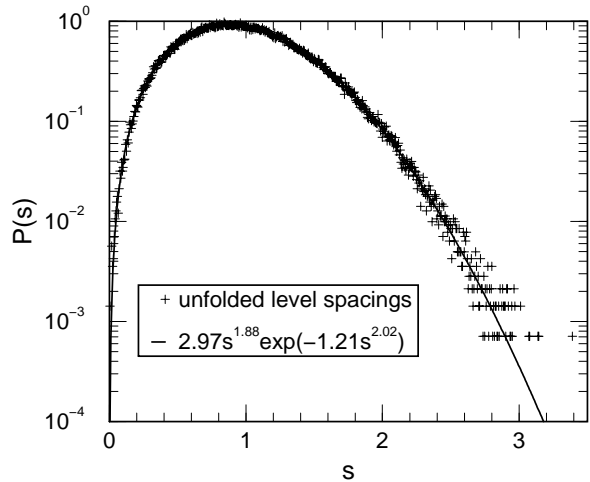}}  
\begin{minipage}[t]{8.1cm}   
\caption{Distribution of the level spacings in the metallic region
at $L/\xi=0.31$. The data ($+$) correspond to $L=C=96$.
The solid line is a fit to the Wigner-surmise form in Eq.~(\ref{ws}).
} 
\label{dists1} 
\end{minipage}   
\end{figure}

In Fig.~\ref{dists1}, the typical $P(s)$ in the quasi-1D metallic
region is shown for $L/\xi=0.31$. The solid line is a fit to the
Wigner-surmise in Eq.~(\ref{ws}) with $\alpha=2.02\pm.02$,
$\beta=1.88\pm0.02$, in very good agreement with
the values expected for the unitary universality class. 
The small discrepancy of $\beta$ from $2$ may be attributed
to insufficient statistics at small $s$ and finite size effects.

\begin{figure}   
\center   
\centerline{\epsfysize=2.8in   
\epsfbox{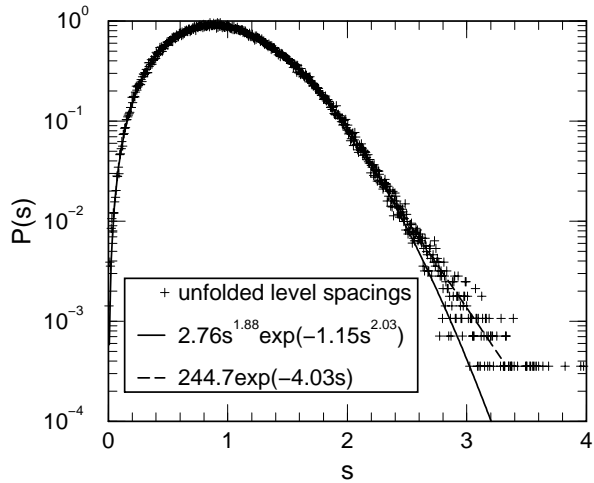}}   
\begin{minipage}[t]{8.1cm}   
\caption{
Distribution of the level spacings in the crossover region
at $L/\xi=1$. The data ($+$) corresponds to $L=C=96$. 
The coexistence of metallic-like (solid line) and
Poisson statistics (dashed line) at large $s$ is evident.
} 
\label{dists2}
\end{minipage}   
\end{figure}

The level statistics becomes more intricate in the crossover regime.
In Fig.~\ref{dists2}, we show $P(s)$ at $L/\xi=1$. The most
interesting feature is the coexistence of metallic-like and insulating-like
statistics. The Wigner-surmise in Eq.~(\ref{ws}),
with $\alpha=2.03\pm0.02$ and $\beta=1.88\pm0.02$,
remains a rather good description (solid-line) of the data 
so long as $s$ is not too large.
On the other hand, the tail at large $s$ is clearly described by
Poisson statistics as shown by the dotted line.
The two different statistics merge together around $s\sim2$.
This kind of hybrid of metallic and insulating behaviors
was first pointed out by Shklovskii, {\it et. al.} to describe
the level statistics at the 3D Anderson metal-insulator transition 
\cite{shk,kramer}.
These results suggest that, in the crossover between the metallic
and insulating regimes, the level correlation becomes finite-ranged,
which manifests itself in the crossover of $P(s)$
from the correlated Wigner-surmise statistics to 
the uncorrelated Poisson statistics at large $s$.
Interestingly, such behaviors are
also observed at the integer quantum Hall transitions \cite{jw}, suggesting
that the results of Shklovskii, {\it et. al.} may be valid for
all delocalization transitions.
Analogous to the conductance distribution $P(g)$ discussed
above, this is likely to be a general property of the {\it critical} 
eigenstates.

\begin{figure}   
\center   
\centerline{\epsfysize=2.8in   
\epsfbox{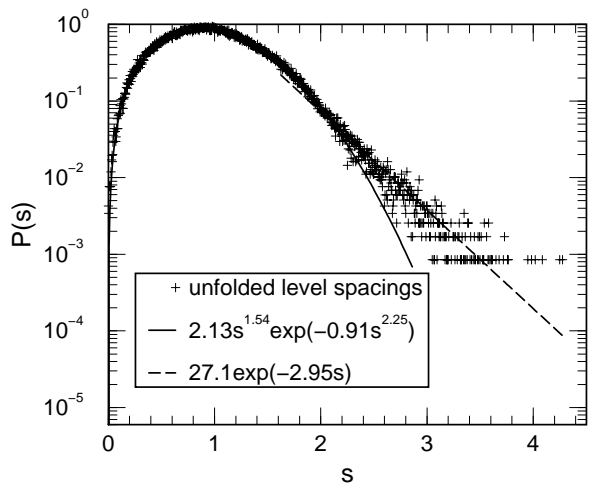}}   
\begin{minipage}[t]{8.1cm}   
\caption{
Distribution of the level spacings in the insulating regime 
at $L/\xi=6.07$. The data ($+$) corresponds to $L=C=96$.  
} 
\label{dists3}  
\end{minipage}   
\end{figure}

The distribution of the level spacings in the insulating regime is 
shown in  Fig.~\ref{dists3} for $L/\xi=6.07$.
As expected, the region of the uncorrelated Poisson tail at large $s$ expands
as a result of the reduction of the correlation range of the levels.
This can be seen by comparing Fig.~\ref{dists2} with Fig.~\ref{dists3}
at large $s$.
One expects $P(s)$ to eventually follow the Poisson distribution
deep in the insulating regime, {\it i.e.} for $L/\xi\gg1$.
It is interesting to point out that the coexistence of metallic
and insulating-like statistics is very robust and extends over
a wide range of $L/\xi$ values.
%
\section{Summary and Discussions}

In conclusion, we have studied the transport properties of the
chiral edge states on the surface of a multilayer integer quantum
Hall state in the field direction. We emphasized the criticality
of the surface state, and that different
behaviors of coherent transport can be reached depending on the path
along which the thermodynamic limit is taken.
The sample aspect ratio is an important quantity that enters the
scaling function of the conductance and conductance fluctuations.
We presented in detail a new and stable algorithm for large scale
conductance calculations in the transfer matrix approach to the 
Landauer formulation. This algorithm allowed, for the first time,
a FFS analysis of the conductance and conductance fluctuations
in the thermodynamic limit of the DN model of the chiral surface
state. The transport properties, in the thermodynamic 
limit approached with a fixed aspect ratio, resemble that of a
quasi-1D conductor, showing a smooth crossover between metallic
and insulating behaviors as a function of the interlayer tunneling.
The asymptotic scaling function of conductance and its variance are found to
be in remarkable agreement with the analytical functions obtained
using the supersymmetric nonlinear $\sigma$-model \cite{mirlin}.
The statistics of the two-terminal conductance
is found to follow the normal distribution in the metallic
and log-normal distribution in the insulating regime.
In the crossover regime, the conductance is broadly
distributed between zero and $e^2/h$, which is well described by
a highly skewed log-normal distribution, similar to
that found at the 2D quantum Hall transitions \cite{wjl}.
We also presented, for the first time, a detailed study
of the level statistics in the eigenvalue spectrum of the
transfer matrix. While the latter in the metallic regime
follows the Wigner-surmise in the unitary universality class,
coexistence of correlated metallic and uncorrelated Poisson statistics
characteristic of an insulating state is found to describe
the data in the crossover regime. We interpret the latter
as a manifestation of the finite range nature
of the level correlations, 
which emerges in the crossover to the insulating regime.

Finally, we briefly discuss the transport behaviors in the 2D metallic
regime (see Fig.~\ref{crossover}). As discussed below Eq.~(\ref{dgscaling}),
in the thermodynamic limit, the 2D chiral metal survives only
in systems with a vanishing aspect ratio.
The behavior of the conductance
and conductance fluctuations can be obtained through FSS of
a sequence of samples with aspect ratio $A\sim1/L^\alpha$ and
$\alpha\ge1$. 
For example, one can approach the asymptotic limit with
$A=a/L$, where $a$ is a constant.
Eqs~(\ref{l0}) and (\ref{xi}) would then lead to
$L/L_0=\sqrt{a/\sigma}$ and $L/\xi=a/2\sigma L$.
The scaling equations for the conductance and conductance fluctuations
in Eqs~(\ref{gscaling}) and (\ref{dgscaling}) become, in the 
asymptotic limit,
\eqa
<g>&=&{e^2\over h}\Gamma\left(0,\sqrt{a\over\sigma}\right), \cr
<\delta g^{2}>&=&\left(e^2\over h\right)^{2n}\Gamma_{2} 
\left(0,\sqrt{a\over\sigma}\right).
\eea
It would be interesting to compare results obtained
this way with the perturbative spin-wave expansion results derived
by mapping the problem to a 1D Su(n,n) quantum ferromagnetic spin 
chain in the limit $n\to0$ \cite{grs}. However, in the DN model,
because the ballistic contribution dominates at large $\sigma$ 
($t^2\to1$), in order to probe the asymptotic 2D regime of vanishing
aspect ratio, one has to go to much larger system sizes \cite{chofisher}.
The stable algorithm present here makes it feasible to carry
out such numerical calculations. These results will be discussed
in a future publication \cite{vwtobe}.

\section{Acknowledgments}

The authors would like to thank Bo\v zidar Jovanovi\'c for many 
useful discussions. They acknowledge useful conversations with
Jan Engelbrecht, Ilya Gruzberg, Dung-Hai Lee, Nick Read, and Subir Sachdev,
and helpful correspondence with Boris Shklovskii.
This work is supported in part
by an award from Research Corporation and the IHRP (500/5011)
at the National High Magnetic Field Laboratory.

\vspace*{\fill}{

}
\end{document}